\documentclass{aa}
\usepackage{graphics, epsfig}
\begin{document}


\title{A new method for the estimate of $H_0$ from quadruply imaged gravitational lens systems}
\author{V. F. Cardone\inst{1}
        \and S. Capozziello\inst{1}
        \and V. Re\inst{1}
    \and E. Piedipalumbo\inst{2}}

\offprints{V.F. Cardone : winny@na.infn.it}

\institute{Dipartimento di Fisica ``E.R. Caianiello'', Universit{\`a} di Salerno and INFN, Sezione di Napoli, Gruppo Collegato di Salerno, Via S. Allende, 84081 - Baronissi
(Salerno), Italy \and Dipartimento di Scienze Fisiche, Universit{\`a} di Napoli and INFN, Sezione di Napoli, Complesso Universitario di Monte S. Angelo, Via Cinthia, Edificio N - 80126 Napoli, Italy}

\date{Receveid / Accepted }


\abstract{We present a new method to estimate the Hubble constant $H_0$ from the measured time delays in quadruply imaged gravitational lens systems. We show how it is possible to get an estimate of $H_0$ without the need to completely reconstruct the lensing potential thus avoiding any {\it a priori} hypothesis on the expression of the galaxy lens model. Our method only needs to assume that the lens potential may be expressed as $r^{\alpha} F(\theta)$, whatever the shape function $F(\theta)$ is, and it is thus able to fully explore the degeneracy in the mass models taking also into account the presence of an external shear. We test the method on simulated cases and show that it does work well in recovering the correct value of the slope $\alpha$ of the radial profile and of the Hubble constant $H_0$. Then, we apply the same method to the real quadruple lenses PG1115+080 and B1422+231 obtaining $H_0 = 58^{+17}_{-15} \ {\rm km \ s^{-1} \ Mpc^{-1}}$\,(68 \% CL). 
}

\titlerunning{A new method for the estimate of $H_0$}

\maketitle

\section{Introduction}
 
Most of the ways of measuring the Hubble constant $H_0$ involve a form of distance ladder, which utilizes a number of astrophysical standard candles and standard ruler relations, and are calibrated locally by a geometrical technique such as trigonometric or dynamical parallaxes (e.g., \cite{M98}). A few methods involve no distance ladder\,: good examples are (i) inferring the distance of the SNeII from their light curves and spectra by modelling their expanded photosphere (\cite{S92}) and (ii) comparing the $H_0$\,-\,independent angular extent of galaxy clusters to their $H_0$\,-\,dependent depth as deduced by the X\,-\,ray emission and the Sunyaev\,-\,Zeldovich microwave background decrement due to the cluster itself (\cite{HB98}).

But the most promising one step method may be considered the one first proposed by Refsdal in 1964 and became feasible only recently thanks to the now available instrumentation. The principle of the Refsdal method is quite simple. If a QSO is multiply imaged by the gravitational lensing effect of a galaxy along the line of sight, the light rays coming from the different images follow different optical path and thus arrive to the observer with a time delay among each other. It is easy to show that these time delays are proportional to the inverse of the Hubble constant $H_0$ and to a factor which depends only on the lensing potential and the source coordinates. Having measured the time delays and estimated the lens dependent factor by the images configuration, one can then obtain a direct estimate of the Hubble constant avoiding all the problems and possible systematic errors connected to the distance ladders. There are actually more than fifty multiply imaged systems (both double and quadruple) and the number of them with measured time delays is increasing (\cite{S00}) so that the prospects of obtaining an accurate estimate of $H_0$ from gravitational lenses is quite good (\cite{K01}). 

However, there is still a major problem connected to the modelling of the lensing galaxies since there are often different models which predict the same images configuration and the other lensing observables. Thus lens modelling is the major source of uncertainty in the Refsdal method. There are two ways to compensate for our lack of knowledge about the lens galaxy. The first is assuming an exact parametric form for the galaxy model and then determine its parameters by fitting to the images positions and time delays ratios (and, eventually, to the flux ratios). However, it is clear that this approach strongly underestimates the uncertainty connected to the lens modelling. Even if one is able to find a parametric galaxy model which is dynamically possible and which reproduces the image properties with acceptably low $\chi^2$, one still has to aggressively explore all other classes of models to get the true uncertainty on $H_0$. But this is not possible with the parametric approach since there is an unavoidable limit to the number of parameters which can be used to describe the galaxy fixed by the number of available constraints. Thus parametric techniques may explore only the simplest models, i.e. one is restricted to a narrow area in the space of the models. To explore it in a systematic fashion one has to follow the second approach using a representation of the galaxy which is as general as possible and thus not restricted to a particular form. One possibility in this sense is to pixelate the galaxy map and consider each pixel as an independent mass element. This is what is done in the {\it pixellated lens method} (\cite{SW97}; \cite{WS00}) which indeed is a valide alternative to the usual parametric techniques. Introducing as less as possible constraints on the reconstructed mass distributions, the pixellated lens method is very efficient in exploring the models space, but it has also the risk of overestimating the uncertainties on $H_0$ since one has almost no way to control if the reconstructed models are physically meaningfull or not. To overcome this difficulty we have elaborated a new approach which aims at estimating the Hubble constant from a detailed exploration of that region of the models space which is compatible with the lensing observables and, at the same time, is physically well motivated. To do this we have to lose some generality since we introduce a lensing potential which has a well defined radial profile, but we still do no hypotheses on the angular part and take also into account the presence of the external shear. Even if less general than the pixellated lens method, our semianalytical technique may be considered as a compromise between the usual parametric technique and the full non\,-\,parametric approach. We do not introduce any defined galaxy lens model, but we take into account all the models which give rise to a lensing potential of the form $r^{\alpha} F(\theta)$, whatever the shape function $F(\theta)$ is, and finally select only the ones which reproduces the lensing observables and are physically well motivated. Thus our method is still able to carefully explore the models space and, at the same time, it does not introduce any overestimate connected to the inclusion of unphysical models.

The plan of the paper is as follows. In Sect. 2 we do some general considerations on the lens equations and write down the relations which will be used to build the method. This is presented in Sect. 3 where we will show how it is possible to algebrically manage the lens equations to finally get a system which is numerically solvable. Since the set of equations is nonlinear, we have elaborated a simple algorithm that allows us to recover all the solutions and exclude the unphysical ones. The code and the selection criteria are then presented in Sect. 4, while the following Sect. 5 is devoted to the application of the method to simulated cases and to the description of how we extract the Hubble constant estimate from the set of solutions. Having so checked that the method indeed works, we can now apply it to real systems; this is the subject of Sect. 6 where we discuss the real lenses PG1115+080 and B1422+231 and obtain our final estimate of $H_0$. Sect. 7 is then devoted to conclusions.

\section{Lens equations and time delays}
 
Let us begin with some general considerations on the estimate of $H_0$ from the time delays between two different images of the same source. To this aim let us fix a cartesian coordinate system with origin on the lens galaxy centre and $(x, y)$ axes along the main axes of the galaxy itself, positevely oriented towards West and North respectively. Let $(r, \theta)$ be the polar coordinates with the $\theta$ angle measured counterclockwise from North. Thus we have\,:
\begin{displaymath}
x = r \sin{\theta} \ , \ \ \ \ y = r \cos{\theta} \ .
\end{displaymath}
The time delay of a light ray deflected by the galaxy lensing effect is given by (\cite{ZP00})\,:
\begin{displaymath}
\Delta t (r, \theta) = h^{-1} \tau_{100} \left [ \frac{1}{2} r^2 - r r_s \cos{(\theta - \theta_s)} 
+ \right .
\end{displaymath}
\begin{equation}
\ \ \ \ \ \ \ \ \ \ \ \ \ \ \left .
+ \frac{1}{2} r_s^2 - \psi(r, \theta) \right ] 
\label{eq: timedelaygen}
\end{equation}
being $(r, \theta)$ the image position, $(r_s, \theta_s)$ the unknown source position and $\psi(r, \theta)$ the lensing potential. In Eq.(\ref{eq: timedelaygen}), $h$ is the Hubble constant $H_0$ in units of 100\,${\rm km \ s^{-1} \ Mpc^{-1}}$ (i.e. $H_0 = 100 h \ {\rm km \ s^{-1} \ Mpc^{-1}}$), while $\tau_{100}$ is a typical time delay estimated for a given set of cosmological parameters $(\Omega_m, \Omega_{\Lambda}, \Omega_{k})$ assuming $H_0 = 100 \ {\rm km \ s^{-1} \ Mpc^{-1}}$.  The typical time delay is defined as\,:
\begin{equation}
\tau_{100} = \left(\frac{D_{OL} D_{OS}}{D_{LS}}\right) 
\frac{(1+z_L)}{c} 
\label{eq: taucento}
\end{equation}
with the usual meaning for the angular diameter distances $D_{OL}, D_{OS}, D_{LS}$; $z_L$ is the redshift of the lens. 

From Eq.(\ref{eq: timedelaygen}) it is easy to get the time delay between two images with coordinates $(r_i, \theta_i)$, $(r_j, \theta_j)$ respectively\,:

\begin{displaymath}
\Delta t_{ij} = \Delta t_i - \Delta t_j = h^{-1} \tau_{100} \ \times 
\end{displaymath}
\begin{displaymath}
\ \ \ \ \ \ \ \ \ 
\left \{ \frac{1}{2} (r_i^2 - r_j^2) 
- r_s [ r_i \cos{(\theta_i - \theta_s)} - r_j \cos{(\theta_j - \theta_s)} ]  \right .
\end{displaymath}
\begin{equation}
\ \ \ \ \ \ \ \ \ \ \ \ \ \left .
- \psi(r_i, \theta_i) + \psi(r_j, \theta_j) \right \} \ .
\label{eq: deltatijpsi}
\end{equation}

According to the Fermat principle, the images lie at the minima of $\Delta t$, so that the lens equations may be simply obtained minimizing $\Delta t$. We get\,:
\begin{equation}
\frac{\partial}{\partial r}\Delta t = 0 \iff 
r - r_s \cos{(\theta - \theta_s)} = \frac{\partial \psi}{\partial r} \ , 
\label{eq: lenseqa}
\end{equation}
\begin{equation}
\frac{1}{r} \frac{\partial}{\partial \theta}\Delta t = 0 
\iff r_s \sin{(\theta - \theta_s)} = \frac{1}{r}\frac{\partial \psi}{\partial \theta} 
\ . 
\label{eq: lenseqb}
\end{equation}
The lensing potential $\psi(r, \theta)$ may be splitted in two terms and it is usually written as\,:
\begin{equation}
\psi(r, \theta) = \psi_{lens}(r, \theta) + \psi_{shear}(r, \theta) \ .
\label{eq: psigen}
\end{equation} 
 The first one is connected with the galaxy lens models through the 2D Poisson equation\,:
\begin{equation}
\nabla^2 \psi_{lens} = 2 \kappa(r, \theta) 
\label{eq: poisson}
\end{equation}
being $\kappa(r, \theta)$ the convergence, i.e. the adimensional surface mass density defined as (\cite{SEF})\,:
\begin{displaymath}
\kappa(r, \theta) = \frac{\Sigma(r, \theta)}{\Sigma_{crit}}
\end{displaymath}
with $\Sigma_{crit} = (c^2/4\pi G) (D_{OS}/D_{OL}D_{LS})$. It is not unusual to suppose that $\psi_{lens}$ may be factorized (see, e.g., \cite{WMK00}); motivated by this, we will assume that\,:
\begin{equation}
\psi_{lens} = r^{\alpha} F(\theta) \ .
\label{eq: psilens}
\end{equation}
Introducing this expression into Eq.(\ref{eq: poisson}) we get\,:
\begin{equation}
\kappa(r, \theta) = \frac{1}{2} r^{\alpha - 2} \left [
\alpha^2 F(\theta) + \frac{d^2 F(\theta)}{d \theta^2} \right ] \ .
\label{eq: kappa}
\end{equation}
Eq.(\ref{eq: kappa}) tells us that our hypothesis on $\psi_{lens}$ simply means that the mass models we are considering are cuspy power\,-\,law and may be spherical or not according to the expression of the shape function $F(\theta)$; since many galactic models predict a single power\,-\,law surface mass density or may be approximated (at least in the external part which is the one we are mainly interested in) by Eq.(\ref{eq: kappa}), we are confident that our hypothesis on $\psi_{lens}$ is not too severe and is quite enough general to explore the mass models degeneracy.  

It is worthwile to point out that lensing potentials of the kind in Eq.(\ref{eq: psilens}) have yet been taken into consideration by different authors in literature. Witt, Mao \& Keeton (2000) used the same potential as in Eq.(\ref{eq: psilens}) to analyse the effect of changing the slope $\alpha$ of the radial profile and the shear strength $\gamma$ (see later) on the estimate of the Hubble constant from the time delays measurements without doing any hypotheses on the shape function $F(\theta)$. On the other hand, introducing a general but parametric expression for $F(\theta)$, Zhao \& Pronk (2001) were able to develop a semianalytical technique to reconstruct the lensing potential in quadruply imaged systems having assigned the boxiness parameter. Their work has been then generalized by Cardone et al. (2001) which have implemented another semianalytical method which is able to recover all the lensing potential parameters without assigning {\it ab initio} any of them. More recently, Evans \& Witt (2001; see also \cite{HW01}) have studied the gravitational lensing properties (i.e. number of images and flux ratios) of the family of scale\,-\,free galaxies with flat rotation curves. As the authors note, the lensing potential of these models are of the form in Eq.(\ref{eq: psilens}) with $\alpha = 1$ and $F(\theta)$ given. Finally, we also note that there are other popular parametric lens models which may be reduced to Eq.(\ref{eq: psilens}). As an example, we remember the isothermal ellipsoidal model (\cite{BK87}) which is obtanined from Eq.(\ref{eq: psilens}) by putting\,:

\begin{displaymath}
\alpha = 1 \ ; F(\theta) \propto \sqrt{\cos^{2}{\theta} + q^{-2} \sin^{2}{\theta}} \ .
\end{displaymath}

These considerations lead us to be confident that our assumption on $\psi_{lens}$ is quite general and allow us to explore in detail a wide area in the space of models.

Let us turn now to the second term in Eq.(\ref{eq: psigen}). This is the potential of the external shear which is introduced to take into account (to the lowest order) of the presence of other galaxies of the group which the lens galaxy belongs to. We have\,:
\begin{equation}
\psi_{shear} = - \frac{1}{2} \gamma r^2 \cos{(2 \theta - 2 \theta_{\gamma})} 
\label{eq: psishear}
\end{equation} 
being $(\gamma, \theta_{\gamma})$ the strength and the position angle (i.e. the angle between the direction of  the shear and the main axis of the lens galaxy) of the shear itself treated as a pseudo\,-\,vector.  

Introducing Eqs.(\ref{eq: psilens}) and (\ref{eq: psishear}) into Eqs.(\ref{eq: lenseqa}) and (\ref{eq: lenseqb}) the lens equations become\,:
\begin{equation}
r - r_s \cos{(\theta - \theta_s)} = \alpha r^{\alpha - 1} F(\theta) 
- \gamma r \cos{(2 \theta - 2 \theta_{\gamma})} \ ,
\label{eq: lensoura}
\end{equation}
\begin{equation}
r_s \sin{(\theta - \theta_s)} = r^{\alpha - 1} f_1(\theta) F(\theta) 
+ \gamma r \sin{(2 \theta - 2 \theta_{\gamma})} 
\label{eq: lensourb}
\end{equation}
having defined\,:
\begin{equation}
f_1(\theta) = \frac{1}{F(\theta)} \frac{dF(\theta)}{d\theta} \ .
\label{eq: funodef}
\end{equation}
Note that this latter definition implicitly assumes that the shape function $F(\theta)$ never vanishes in the positions of the images. Thus the method we develop will work only for those potentials satisfying this constraint; however, this is not a seriuos problem as it may be shown analysing many of the potentials in literature. We will also assume that $f_1(\theta)$ never vanishes in the positions of the images; this constrains us to consider only not spherically symmetric potentials since for the spherical potentials $F(\theta)$ is constant and thus $f_1(\theta)$ identicaly vanishes. On the other hand, lensing galaxies are usually elliptical ones and there are also different evidences that dark halos are not spherical (see, e.g.,\cite{S99} and references therein). So these considerations suggest us that to consider only not spherically simmetric potentials is not a serious limitation. It is useful to solve Eq.(\ref{eq: lensoura}) with respect to $\psi_{lens}$; we simply get\,:
\begin{displaymath}
\psi_{lens}(r_i, \theta_i) = r_i^{\alpha} F(\theta_i) =
\end{displaymath}
\begin{equation}
\ \ \ \ \ \ \ \ \ \ \ \ \ \ \  
\frac{r_i^2 - r_i r_s \cos{(\theta_i - \theta_s)} 
+ \gamma r_i^2 \cos{(2 \theta_i - 2\theta_{\gamma})}}{\alpha}
\label{eq: ftheta}
\end{equation}
where hereinafter we add a label (running from $i$ to $l$) to the image coordinates to distinguish among them. Note that this relation diverges if $\alpha = 0$, but this hypothesis may be excluded since $\alpha = 0$ means that $\psi_{lens}$ does not depend on $r$ which is very unlikely. Introducing Eqs. (\ref{eq: psishear}) and (\ref{eq: ftheta}) into Eq.(\ref{eq: psigen}) and the result into Eq.(\ref{eq: timedelaygen}), it is then simple to show that the time delay between two images $i$ and $j$ is\,:
\begin{displaymath}
\Delta t_{ij} = \Delta t(r_i, \theta_i) - \Delta t(r_j, \theta_j) =  
\Delta t_i - \Delta t_j =
\end{displaymath}
\begin{displaymath}
\ \ \ \ \ \ \ \ 
\frac{h^{-1} \tau_{100}}{2 \alpha} 
\left \{ (\alpha - 2) (r_i^2 - r_j^2) + 2 (1 - \alpha) r_s \ \times \right .
\end{displaymath}
\begin{displaymath}
\ \ \ \ \ \ \ \ \
\left [
r_i \cos{(\theta_i - \theta_s)} - r_j \cos{(\theta_j - \theta_s)} \right ] + 
(\alpha - 2) \ \times
\end{displaymath}
\begin{equation}
\ \ \ \ \ \ \ \ \left .
\left [
r_i^2 \cos{(2 \theta_i - 2 \theta_{\gamma})} 
- r_j^2 \cos{(2 \theta_j - 2 \theta_{\gamma})} \right ] \right \} \ .
\label{eq: delay}
\end{equation}
Eq.(\ref{eq: delay}) shows us that to estimate the Hubble constant $h$ from the measured time delay between images $i$ and $j$ one has to know the slope $\alpha$ of the radial profile, the source coordinates $(r_s, \theta_s)$ and the shear parameters $(\gamma, \theta_{\gamma})$. It also shows us that it is not strictly necessary to completely reconstruct the lensing potential $\psi_{lens}$ i.e. we do not need to know anything about the shape function $F(\theta)$ if we were able to find the set of quantities $(\alpha, r_s, \theta_s, \gamma, \theta_{\gamma})$. To this aim one usually adopt a parametric approach giving a lens mass model which depend on some unknown parameters. These latter, the source coordinates and the shear parameters are then found out using a minimization technique which aims at recovering those values of the parameters such that the predicted images positions and time delays agree well with what is observed. This approach is plagued by the contrast between the need to introduce many parameters to build up a reliable and accurate galaxy model and the fixed number of constraints (eigth from the four image positions and two from the time delay ratios). As a consequence the number of degrees of freedom may be too low to get an accurate enough estimate of the parameters. To increase the number of degrees of freedom one may also consider the flux ratios, but this is dangerous since these latter may be contaminated by microlensing (\cite{CR79}, \cite{KdB00}) and other secondary effect, such as, e.g., substructure in the lens galaxy (\cite{MS98}, which are difficult to handle. Beside, there are often different models which fit well the same set of observations so that the parametric approach is unable to fully explore the degeneracy in the lensing potential leading to an underestimate of the systematic errors on $h$. In the next Section we will show how it is possible to overcome this problem and obtain an estimate of the Hubble constant taking into account the degeneracy in the lens mass models.

\section{A new semianalytical method to estimate $H_0$}

Eq.(\ref{eq: delay}) tells us that it is possible to estimate the (dimensionless) Hubble cosntant $h$ having measured the time delays between any pair of images and their positions provided that one is able to get the slope $\alpha$ of the lens potential radial profile, the source coordinates $(r_s, \theta_s)$ and the shear parameters $(\gamma, \theta_{\gamma})$. Note that it is not strictly needed to know the exact expression of the shape function $F(\theta)$ if we were able to recover the above parameters. This suggests that it should be possible to work out a method which leads to an estimate of $h$ whatever is the shape function itself provided that the lensing potential may still be written as in Eq.(\ref{eq: psigen}) with $\psi_{lens}$ as in Eq.(\ref{eq: psilens}). Actually this is true as we will show in this Section. 

As a preliminary consideration, let us observe that the lens equations (\ref{eq: lenseqa}), (\ref{eq: lenseqb}) may be written for each one of the images thus giving us eigth linearly independent equations. Whatever is the way we manipulate these equations we may still obtain the same number of independent equations which may be used to determine (at least in principle) eight unknown variables. In these equations $F(\theta_i)$, $F(\theta_j)$, $F(\theta_k)$ and  $F(\theta_l)$ are not functions, but numbers which we may rename as $(Fi, Fj, Fk, Fl)$; a similar discussion also holds for the values of $f_1(\theta)$. So we have to determine these set of numbers and not the shape function itself. The degeneracy in the lensing potential may be reformulated as follows\,: {\it all the shape functions $F(\theta)$ such that their values in the positions of the images are equal to the set of numbers $(Fi, Fj, Fk, Fl)$ are acceptable since the predicted image positions and time delay ratios are the same as the observed ones}. This observation explains why we do not need to give an {\it a priori} expression for the potential $\psi_{lens}$\,: we do not solve for the potential itself, but only for the values of the potential in the positions of the images. These latter are all what we need to finally get an estimate of the Hubble constant. As we will see, however, we really do not solve with respect to $(Fi, Fj, Fk, Fl)$, but with respect to $(f1i, f1j, f1k, f1l)$, being these latter the values of $f_1(\theta)$ in the position of the images. This is related to the well known circumstance that the image positions are determined by the first derivative of the potential and not directly by the potential itself. Our previous discussion on the lens models degeneracy still holds when considering $f_1(\theta)$. 

Let us now turn to the bulk of the method. As a fisrt step, let us write Eq.(\ref{eq: lensourb}) for the image $i$ and solve it with respect to $f1i$. It is easy to get what follows\,:

\begin{displaymath}
\frac{1}{f1i} = \frac{r_i^{\alpha - 1} Fi}{r_s \sin{(\theta_i - \theta_s)} 
- \gamma r_i \sin{(\theta_i - \theta_{\gamma})}} \ .
\end{displaymath}

Solving Eq.(\ref{eq: lensoura}) with respect to $r_i^{\alpha - 1} Fi$ and inserting this solution in the previous relation we get the following equation\,:

\begin{equation}
\frac{\alpha}{f1i} = \eta_i = 
\frac{r_i - r_s \cos{(\theta_i - \theta_s)} + \gamma r_i \cos{(2\theta_i - 2\theta_{\gamma})}}
{r_s \sin{(\theta_i - \theta_s)} - \gamma r_i \sin{(\theta_i - \theta_{\gamma})}} 
\label{eq: firstgroup}
\end{equation}

where we have introduced a new variable $\eta_i$. Eq.(\ref{eq: firstgroup}) may be rewritten for each one of other three images using the right coordinates. Thus we get a system of four independent equations in eight unknown variables, i.e. the four quantities $(\eta_i, \eta_j, \eta_k, \eta_l)$, the source coordinates $(r_s, \theta_s)$ and the shear parameters $(\gamma, \theta_{\gamma})$. To close the system we need other four independent equations. To this aim, let us turn back to Eq.(\ref{eq: lensourb}) written for image $i$ and solve it with respect to $r_i^{\alpha - 1} Fi$. Inserting the solution into Eq.(\ref{eq: lensoura}) after some simple algebra one finally gets\,:

\begin{displaymath}
[1 + \eta_i \tan{(\theta_i - \theta_s)}] r_s \cos{(\theta_i - \theta_s)} =
\end{displaymath}
\begin{equation}
\{1 + [1 + \eta_i \tan{(2\theta_i - 2\theta_{\gamma})}]\} \gamma r_i \cos{(2\theta_i - 2\theta_{\gamma})} \ .
\label{eq: secondgroup}
\end{equation}

We may write down Eq.(\ref{eq: secondgroup}) also for the other three images thus obtaining a total of four equations of this kind. Adding these relations to the four previously found, we finally have a system of eight linearly independent equations in the eight unknowns $(r_s, \theta_s)$, $(\gamma, \theta_{\gamma})$, $(\eta_i, \eta_j, \eta_k, \eta_l)$ which may be solved (at least numerically). Note that there is still a degeneracy in the system since we solve with respect to $(\eta_i, \eta_j, \eta_k, \eta_l)$ and thus we are not able to get the slope $\alpha$ of the radial profile which is the quantity we are mainly interested in since it appears in the time delay equation (\ref{eq: delay}) together with the source coordinates $(r_s, \theta_s)$ and the shear parameters $(\gamma, \theta_{\gamma})$. But we have still not used all the informations we have at our disposal. Since our aim is to determine the Hubble constant from the time delays, it is obvious that we have measured these quantities and so we also know the time delay ratios. Using Eq.(\ref{eq: delay}) we may write down the time delay ratios as function of $(\alpha, r_s, \theta_s, \gamma, \theta_{\gamma})$. Having yet found out the source coordinates and the shear parameters solving the system, this latter relation may then be solved with respect to $\alpha$. Note that now we have all what we need to estimate $h$ since we can now solve Eq.(\ref{eq: delay}) with respect to $h$ itself being all the other quantities known. However, we may obtain also the values of lensing potential $\psi(r, \theta)$  in the position of the four images. Actually, having found out $\alpha$ and knowing the quantities $(\eta_i, \eta_j, \eta_k, \eta_l)$, we may immediately estimate $(f1i, f1j, f1k, f1l)$, then obtain $(Fi, Fj, Fk, Fl)$ from Eq.(\ref{eq: lensoura}) and finally get\,:

\begin{equation}
\psi(r_i, \theta_i) = r_i^{\alpha} Fi - \frac{1}{2} \gamma r_i^2 \cos{(2\theta_i - \theta_{\gamma})} 
\label{eq: psisolved}
\end{equation}
with\,:
\begin{equation}
Fi = \frac{r_i - r_s \cos{(\theta_i - \theta_s)} + \gamma r_i \cos{(2\theta_i - 2\theta_{\gamma})}}
{\alpha r_i^{\alpha - 1}} \ .
\label{eq: fisolved}
\end{equation}

The method describeed here thus allows us to get an estimate of the Hubble constant $h$ without assigning any {\it a priori} expression for the shape function and it is thus able to fully explore the lens model degneracy whatever is the angular structure of the lens which is crucial in determining the image configurations. Note that to this aim the method does not need to use the flux ratios which may be seriously affected by microlensing and other systematics which may lead to significatively wrong results. Beside, the method also takes into account the effect of the other galaxies of the group which the lens belongs to through the introduction of the external shear whose parameters are also obtained. 

\section{Solving the system of equations and estimating the Hubble constant $H_0$}

To recover the whole set of parameters we need to estimate the Hubble constant from the time delay we use the procedure we briefly describe here. First, we substitute the $\eta_i$ from Eq.(\ref{eq: firstgroup}) into Eq.(\ref{eq: secondgroup}) and do this for each image finally obtaining a system of only four equations in the source coordinates $(r_s, \theta_s)$ and $(\gamma, \theta_{\gamma})$ that we do not write here for sake of shortness. Then, supposing to have solved this system, we estimate $\alpha$ from the time delay ratios and use this estimate and Eqs.(\ref{eq: secondgroup}) to get $(f1i, f1j, gf1k, f1l)$. Finally, we evaluate $(Fi, Fj, Fk, Fl)$ from Eq.(\ref{eq: fisolved}) and then the Hubble constant from the time delay with the values of the lensing potential given by Eq.(\ref{eq: psisolved}). Note that the most problematic step in this procedure is solving the system of four equations in $(r_s, \theta_s)$ and $(\gamma, \theta_{\gamma})$ since this is a nonlinear set of equations and its solutions may be found only numerically. This has led us to develop an algorithm\footnote{The code is nothing more but a notebook written for MATHEMATICA, which we have named {\it HERQuLeS} ({\it $H_0$ Estimate Recovered from Quadruple Lens Systems}).} to search for the solutions of the system. To avoid introducing any bias in the search, we give to the algorithm $\cal{N}$ random starting points for $(r_s, \theta_s, \gamma, \theta_{\gamma})$, where $\cal{N}$ is a number fixed by the user\footnote{$\cal{N}$ should be large to explore a wide region in the parameter space, but not too large to save computer time. The right choice must be a compromise between these two different circumstances. A possible strategy could be to fix $\cal{N} =$\,10000 and then, eventually, run HERQuLeS more than one time if necessary.}. We then obtain $\cal{N}$ solutions which are not all physically acceptable. To select among these we have imposed a set of selection criteria\,: the code checks the list of $\cal{N}$ solutions and finally retains only the ones satisfying the whole set of criteria. Schematically the selection criteria we impose and the reasons why we use them are described in the following.

\begin{figure*}[ht]
\centering
\resizebox{10cm}{!}{\includegraphics{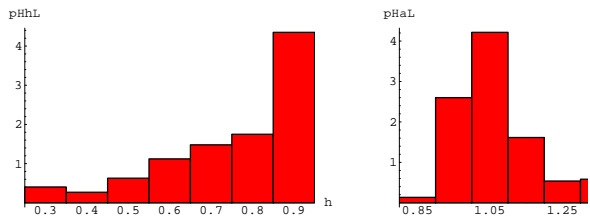}}
\hfill
\caption{Histograms of the recovered values of the slope $\alpha$ (right) and of the Hubble constant $h$ (left) for simulation 1 normalized such that the total area under each histogram is one.}
\end{figure*}

\begin{enumerate}

\item{$0 < r_s < \max{(r_i, r_j, r_k, r_l)}$\,: we impose this cut on $r_s$ since it is not possible that the source is outside the {\it ring} delineated by the more distant of the images\footnote{A nice demonstration of this may be obtained using the web tool developped by K. Ratnatunga which generates the image of a lens system given the lensing potential, the source coordinates and the observational characteristics (see {\tt http://mds.phys.cmu.edu/ego\_cgi.html}).}.}

\item{$0 < \alpha < 2$\,: from Eq.(\ref{eq: kappa}) it is evident that the surface mass density scales as $r^{\alpha - 2}$ so that $\alpha < 2$ is needed in order to have $\kappa(r, \theta)$ monotonically decreasing with the radius; at the same time, it may be shown that the projected mass inside $r$ goes as $r^{\alpha}$ so that $\alpha$ must be positive to be physically reasonable.}

\item{$0 < \gamma < \gamma_{max}$\,: the shear intensity $\gamma$ is positive by definition and is not expected to be too large since it is a perturbation; for this reason we impose an upper limit $\gamma_{max}$ which may be fixed quite arbitrarily; we decide to use $\gamma_{max} = 0.3$, as in Witt et al. (2000), but another choice could be to impose $\gamma = \gamma_{crit}$ (being this latter defined such that for values of $\gamma \ge \gamma_{crit}$ the estimated $H_0$ becomes negative) which depends on the particular lens system to be considered (O. Wucknitz, priv. comm.).}

\item{{\it the shear is approximately well directed}\,: by this we mean that the position angle $\theta_{\gamma}$ of the shear must be directed towards the supposed origin of the shear itself; e.g., if the shear were due to the cluster which the lens galaxy belongs to, then $\theta_{\gamma}$ should be aligned with the cluster mass center since there are no reasons why it should point elsewhere.}

\item{{\it $(Fi, Fj, Fk, Fl)$ are all positive}\,: this constraint turns out from the consideration that the lensing potential $\psi_{lens}(r, \theta)$ is positive defined; since $r^{\alpha}$ is always positive, so must be the quantities $(Fi, Fj, Fk, Fl)$ in order to have a lensing potential physically acceptable.}

\item{{\it the set of parameters so found solve the lens equations}\,: this simply means that we insert the solutions in the lens equations (\ref{eq: lensoura}), (\ref{eq: lensourb}) to check if the solutions found is a correct one or the result of a wrong convergence of the numerical algorithm.}

\item{$h_{min} \le h \le h_{max}$\,: this condition is imposed to avoid that the recovered estimate of $h$ is not physical; in principle, one should fix $(h_{min}, h_{max}) = (0, 1)$ to not introduce any bias in the estimate of $h$, but it is also well known that different method of estimates never give values of $h$ less than 0.3 so that we have used $(h_{min}, h_{max}) = (0.25, 0.95)$.}

\end{enumerate}

Obviously, one may also change the order of the constraints; we have chosen this one as a tentative to minimize the CPU time needed, but nothing prevents to add or remove some selection criteria. At the end of the selection procedure, one has a set of $\cal{M}$ solutions each one with attached an estimate of the Hubble constant $h$. Note that these solutions could also be different from each other (and actually this is what happens) since the system we have initially solved is nonlinear and thus may have more than one solution. This is another way to see the lens models degeneracy. A final estimate of $h$ may be obtained considering the mean as central value and as acceptable range the one which contains the 90\% (or the 68\%) of the values so found thus finding out what we will call the 90\% CL (or 68\% CL) range. As we will see in the next section, this will give a quite large range, but this is simply a consequence of having taken fully into account the lens model degeneracy. However, we will also see how it is possible to narrow the range for $h$ combining the data from different lens systems.

\section{Application to simulated systems : refining the estimate of $H_0$}

To test if our method indeed works and to see whether it is possible to reduce the uncertainty on the estimate of the Hubble constant $H_0$ we have constructed three simulated quadruple systems. To this aim, we have used the following expression for the shape function\,:

\begin{equation}
F(\theta) = | 1 - \delta \cos{(2\theta - 2\theta_p)} |^{\beta}
\label{eq: shapesimul}
\end{equation}
where $\beta$ is a boxiness parameter, $\delta$ is a flattening indicator and $\theta_p$ is the position angle of the lensing potential (see \cite{ZP00} and \cite{Nostro} for further details). Different choices of the three parameters $(\beta, \delta, \theta_p)$ and of the slope $\alpha$ of the radial profile allows to generate different simulated systems provided that one have also fixed both the source coordinates $(r_s, \theta_s)$ and the shear parameters $(\gamma, \theta_{\gamma})$. To compute the time delays one has also to choose the cosmological parameters $(\Omega_m, \Omega_{\Lambda}, \Omega_k, h)$ and the redhsifts of lens ($z_L$) and source ($z_S$). We adopt a flat cosmological scenario for all our simulations fixing\,:

\begin{displaymath}
(\Omega_m, \Omega_{\Lambda}, \Omega_k, h) = (0.3, 0.7, 0.0, 0.72) 
\end{displaymath}
and\,:
\begin{displaymath}
(z_L, z_S) = (0.310, 1.722) \longrightarrow \tau_{100} = 33.37 \ {\rm days \ arcsec^{-2}} \ .
\end{displaymath}
The other parameters were fixed as resumed in the following Table 1.

\begin{table}[h]
\begin{center}
\begin{tabular}{|c|c|c|c|} 
\hline
Id & $(r_s, \theta_s)$ & $(\gamma, \theta_{\gamma})$ & $(\alpha, \beta, \delta, \theta_p)$ \\
\hline
S1 & $(0.09, 40^o)$ & $(0.20, 36^o)$ & $(1.0, 1.0, 0.07, 45^o)$ \\
S2 & $(0.09, 40^o)$ & $(0.12, 0.36)$ & $(1.2, 1.0, 0.13, 45^o)$ \\
S3 & $(0.10, 15^o)$ & $(0.15, 66^o)$ & $(1.1, -0.25, -0.15, 32^o)$ \\
\hline
\end{tabular}
\end{center}
\caption{Source coordinates, shear and lensing potential parameters for the simulated lenses.}
\end{table} 

\begin{figure*}[ht]
\centering
\resizebox{10cm}{!}{\includegraphics{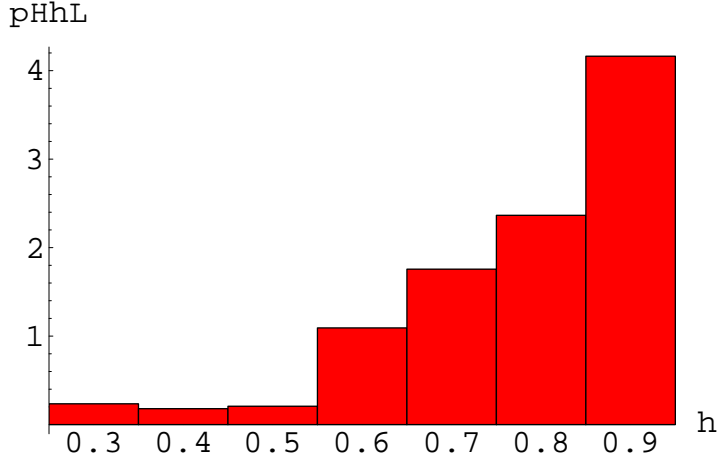}}
\hfill
\caption{Final histogram (normalized such that the total area is one) of the recovered values of the Hubble constant $h$ obtained combining the results from the three simulations as described in the text.}
\end{figure*}

The results of applying the method to these simulated systems is well resumed in Fig. 1 where, as an example, we plot the histograms of the recovered values of the slope $\alpha$ and of the Hubble constant $h$ for our simulation 1. The first interesting result is that the method indeed works well in recovering the $\alpha$ parameter which is, perhaps, the most important. Considering the 90\% range, we get\,:

\begin{displaymath}
\alpha = 1.04^{+0.23}_{-0.15} \ \ for \ simulation \ 1 \ ; 
\end{displaymath}

\begin{displaymath}
\alpha = 1.18^{+0.29}_{-0.21} \ \ for \ simulation \ 2 \ ; 
\end{displaymath}

\begin{displaymath}
\alpha = 1.13^{+0.25}_{-0.17} \ \ for \ simulation \ 3 \ .
\end{displaymath}

Comparing these results with the values in Table 1, one sees that the method recovers the correct value of $\alpha$ with a good enough accuracy since the central value of the delineated range for each simulation is very close to the input one. The uncertainty is not too large ($\sim 20 \%$) which will impact mostly on the estimate of $h$. Actually, from Eq.(\ref{eq: delay}) it is clear that $\alpha$ and $h$ are anticorrelated so that a little error on $\alpha$ leads to a large uncertainty on $h$. This is clearly shown in the histograms for $h$ which lead to the following estimates (90\% CL)\,:

\begin{displaymath}
h = 0.76^{+0.19}_{-0.26} \ \ for \ simulation \ 1 \ ; 
\end{displaymath}
 
\begin{displaymath}
h = 0.74^{+0.21}_{-0.32} \ \ for \ simulation \ 2 \ ; 
\end{displaymath} 

\begin{displaymath}
h = 0.73^{+0.22}_{-0.30} \ \ for \ simulation \ 3 \ . 
\end{displaymath}

\begin{figure*}[ht]
\centering
\resizebox{10cm}{!}{\includegraphics{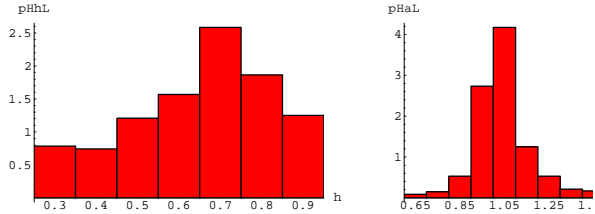}}
\hfill
\caption{Histograms of the recovered values of the slope $\alpha$ (right) and of the Hubble constant $h$ (left) for PG1115+080 normalized such that the total area under each histogram is one.}
\end{figure*}

The central value of the range is close to the input $h$, but the uncertainty is quite large ($\sim 40\%$). However, if not encouraging, this result is not unexpected because of the yet discussed anticorrelation between $\alpha$ and $h$. It is once again a consequence of having taken fully into account the lens model degeneracy and so it is not a surprise that our estimates have a so large uncertainty. On the other hand, however, we may try to reduce this uncertainty combining the results from different lens systems since it is as we were marginalizing on the model parameters to retain only the Hubble constant. To combine different results we proceed as follows (see also \cite{WS00}). For each simulation, we divide the set of values recovered for $h$ in bin of 0.01 and to each bin we assign a probability defined as the number of values in that bin divided by the total number of results. Then we build a combined histogram giving to each bin a probability which is equal to the product of the three probabilities from each simulation. Finally, we delete from the sample those bins which have a combined probability less than 1\% since these values turn out to be very unlikely. Note that this procedure reduces the number of values in the final sample, but we believe that this does not lead to exclude any possibly important value since the excluded bins are really unlikely. The result is the histogram in Fig. 2, where we plot bins of width 0.1 for sake of clarity. The combined histogram leads us to the following estimates for the Hubble constant\,:

\begin{displaymath}
H_0 = 78 \pm 13 \ {\rm km \ s^{-1} \ Mpc^{-1}} \ \ at \ the \ 68 \ \% \ CL \ ;
\end{displaymath}

\begin{displaymath}
H_0 = 78_{-19}^{+17} \ {\rm km \ s^{-1} \ Mpc^{-1}} \ \ at \ the \ 90 \ \% \ CL \ .
\end{displaymath}

The uncertainty has been indeed reduced to $\sim 20\%$ which is a nice result and suggests that it is possible to further reduce the range adding other systems.

It is worthwile to spend some words about the shape of the histograms in Figs. 1 and 2. These could suggest an estimate of $h$ near $\sim 0.9$ where the probability gets higher, whilst we report $h \sim 0.7 \div 0.8$ as best estimate. This is due to having chosen to use the arithmetic mean as best estimate; even if the last bin has the higher probability, this latter is less than the sum of the remaining ones and this cause the mean to be lower than what Figs. 1 and 2 could suggest. If we had chosen to consider as best estimate of $h$ the value corresponding to the bin with the highest probability, we should get an higher value for $h$ in disagreement with the input one. This could be explained if one considers that it is the full ensemble of models which must be take into account when determining the Hubble constant and not only the most likely ones. To use the mean of the sample as estimate of $h$ is a simple way to fully taken into account the lensing model degeneracy. Our results from the simulations confirm this expectation and further enforce our adopted procedure for the estimate of the Hubble constant.

We do not discuss here the results obtained for the shear parameters and the source coordinates since these quantities are less interesting. We limit ourselves to say that the central values recovered by the method are always close to the input ones, but the uncertainties turn out to be quite large.

\begin{figure*}[ht]
\centering
\resizebox{10cm}{!}{\includegraphics{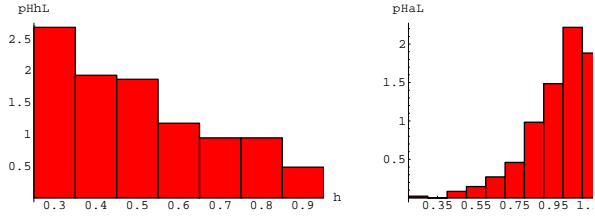}}
\hfill
\caption{Histograms of the recovered values of the slope $\alpha$ (right) and of the Hubble constant $h$ (left) for B1422+231 normalized such that the total area under each histogram is one.}
\end{figure*}

\section{Application to real systems : PG1115+080 and B1422+231}\

The simulations described in the previous section have shown that the method indeed works recovering the correct values both of the slope $\alpha$ of the radial profile and of the Hubble constant $H_0$. Given these encouraging results, now we apply the method to real quadruply imaged systems. To this aim, we need a system with four images and a good astrometry not only of the images position, but also of the lens galaxy centre since this latter will be used as the origin of our coordinate system. Beside, we also need that there is only one galaxy acting as lens otherwise it is not possible to factorize the lensing potential. Finally, to estimate the value of the Hubble constant we need that the time delays (or, at least, one time delay and the ratio between any pair of delays) have been measured. Searching the CASTLES database (\cite {CASTLES}), we have found that there are only two systems which satisfy these requirements\footnote{There is also a third lens system which may seem to be interesting, i.e. B1608+656. Actually, this system may not be considered since there are two lensing galaxies (probably undergoing a merging event) so that the term $\psi_{lens}$ in the lensing potential is not separable and thus our method cannot be applied.}\,: PG1115+080 and B1422+231. In what follows, we first apply the method to the two systems separately and then combine the results to get a better estimate of the Hubble constant. In all the applications we adopt a flat cosmology with $(\Omega_m, \Omega_{\Lambda}) = (0.3, 0.7)$; the effect of changing these cosmological parameters is of the order of few percent so that we may neglect it.

\subsection{Application to PG1115+080}
 
The first system we consider is the well known PG1115+080, first discovered by Weymann et al. (1980) and then studied in detail by several authors (see, e.g, \cite{KK97}; \cite{WS00}; \cite{ZP00}; \cite{Nostro}) with different techniques. This system consists of four images (named $A1$, $A2$, $B$ and $C$) of a radio quiet QSO at $z_S = 1.722$, while the lens is an elliptical galaxy belonging to a group of $\sim$\,10 galaxies at $z_L = 0.310$. The center of the group is at $(r_g, \theta_g) = (20'' \pm 0.2'', -117^o \pm 3^o)$ and its effect has been taken into account as an external shear in previous models. We have applied the method to this system using as images coordinates the ones measured by Impey et al. (1998) with HST observations; the time delay between images $A1$ and $A2$ is too small to be measured, while the one between images $B$ and $C$ is $\Delta t_{BC} = 25.0 \pm 1.7$\,d (\cite{Barkana}) with image $B$ arriving last. There are also two different measures of the time delay ratio among images $BC$ and $AB$\,: Schechter et al. (1997) first reported $r_{ABC} = \Delta t_{AB}/\Delta t_{BC} = 0.7 \pm 0.3$, while a later analysis by Barkana (1997) found $r_{ABC} = 1.13 \pm  0.18$. It is this last value which we consider more reliable and use in the application of the method. Fig. 3 shows the histograms of the recovered values of the slope $\alpha$ of the radial profile and of the dimensionless Hubble constant $h$. Our main results (given as 90\% CL) are the following ones\,:

\begin{equation}
\alpha = 1.03_{-0.20}^{+0.24} \ \ \ \ ; \ \ \ \ h = 0.68_{-0.32}^{+0.25} \ \ \ \ .
\label{eq: estpg}
\end{equation} 

The slope $\alpha$ of the radial profile is practically equal to the one of the isothermal model, consistent with what is expected if galaxies are embedded in a dominant isothermal dark halo. The central value for $h$ is also very near to the previous estimate of the Hubble constant obtained for this system. PG115+080 has been studied in detail by different authors using different techniques. Keeton \& Kochanek (1997) have used the usual least $\chi^2$ parametric approach modelling the lensing potential as the sum of a term due to the galaxy and an external shear from the group. They have used various elliptical models and have found that none of them may fit the image positions without any external shear. A carefull analysis of the possible systematic effects and of the different models also including the external shear from the group finally lead them to estimate the Hubble constant as $H_0 = 51_{-13}^{+14} \ {\rm km \ s^{-1} \ Mpc^{-1}}$. Our estimated 90\% range does overlap well with their one which is a very encouraging result since our method is completely different.  We also note that they also considered the isothermal model which turns out to fit well the lensing observables. Unfortunately their isothermal model is a spherical one and thus its lensing potential does not fall in the class of the models considered by our method. However, the same estimate of $H_0$ is obtained even if one sligthly flattens their model. The result is the isothermal ellipsoidal model (\cite{BK87}) which belongs to the class of potentials (\ref{eq: psilens}), as we have yet noted in Sect. 2, and also fits well the image positions. Zhao \& Pronk have applied their semianalitycal method to PG1115+080 using a lensing potential which is obtained from Eq.(\ref{eq: psilens}) fixing $F(\theta)$ as in Eq.(\ref{eq: shapesimul}). They examine two different models with fixed valued of the boxiness parameter $\beta$ and finally turns out with two quite different estimates of $H_0$ ranging from 20 to 50\,${\rm km \ s^{-1} \ Mpc^{-1}}$ or from 50 to 90\,${\rm km \ s^{-1} \ Mpc^{-1}}$. The first case is marginally consistent with our result, whilst the second one is in good agreement. A direct comparison with the value of the slope $\alpha$ is not possible since they do not report the estimate for this latter parameter, but only three vales for what they call the {\it effective power\,-\,law slope}. In an our previous paper (\cite{Nostro}) we have presented a semianalytical technique to reconstruct a lensing potential as in Eq.(\ref{eq: psilens}) with $F(\theta)$ as in (\ref{eq: shapesimul}) using as constraints only the images positions. Then we have applied this method to PG1115+080 obtaining $H_0 = 56_{-11}^{+12} \ {\rm km \ s^{-1} \ Mpc^{-1}}$ still in agreement with the present result in Eq.(\ref{eq: estpg}). It is also enncouraging that in that paper we get $\alpha = 1.12$ still in agreement with the estimate (\ref{eq: estpg}) obtained with the new approach used here. Another interesting comparison may be done with the results obtained on PG1115+080 using the {\it pixellated lens method} (\cite{WS00}) since this is a not parametric technique. Their 90\% CL estimate ranges from 30 to 75\,${\rm km \ s^{-1} \ Mpc^{-1}}$ which well overlaps our estimated range. Note also that the uncertainties on $h$ from the pixellated method are of the same order as our own reflecting once again the fact that both methods fully take into account the lens models degeneracy. These successfull comparisons confirm lead us to be quite confident in the validity of our new method since it turns out to give results in good agreement with all the previous ones in literature obtained by using different technique but lensing potentials belonging to the same class we consider here. Finally we also note that the uncertainty on our estimates of $\alpha$ and the wide 90\% CL for $h$ are of the same order as the ones we have obtained from the simulations thus leading us to consider another lens system to reduce the estimated range for the Hubble constant.

\begin{figure*}[ht]
\centering
\resizebox{10cm}{!}{\includegraphics{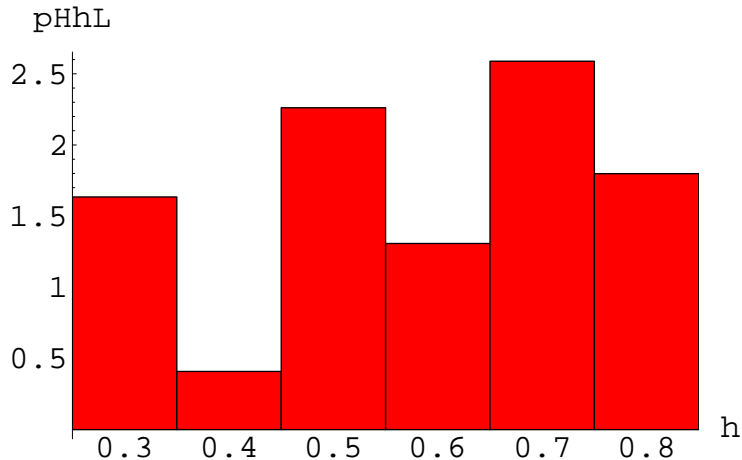}}
\hfill
\caption{Final histogram (normalized such that the total area is one) of the recovered values of the Hubble constant $h$ obtained combining the results from PG1115+080 and B1422+231 as described in the text.}
\end{figure*}

\subsection{Application to B1422+231}

The second system we consider is B1422+231 fisrt discovered by Patnaik et al. (1992) and then observed in detail both in optical (\cite{L92}; \cite{Ietal96}) and in radio (\cite{Petal99}). The images configuration is quite different from the typical {\it cross} which is observed in quadruply systems (e.g. PG1115+080 and G2237+030) since we have three images on a side of the main lens galaxy and a fourth one very far on the other side. This suggest that the source is located near a cusp and is not aligned with the galaxy centre as often in quadruply systems. The four images have redshift $z_S = 3.62$ while the main lensing galaxy is at $z_L = 0.49$. There are other two galaxies not too far from the main one at the same redshift; we will consider them as the sources of the external shear and impose that our reconstructed shear points in the direction between these two galaxies. As input we use the images and galaxies positions determined by Impey et al. (1996) with HST observations, while we take as time delays the recently determined values by Patnaik \& Narasimha (2001) even if these latter have a very large uncertainty. Fig. 4 shows the histograms for $\alpha$ and $h$ obtained applying our method. The main results we get are\,:

\begin{equation}
\alpha = 1.08_{-0.36}^{+0.33} \ \ \ \ ; \ \ \ \ h = 0.47_{-0.22}^{+0.31} \ \ \ \ .
\label{eq: estb}
\end{equation} 

The slope $\alpha$ of the radial profile is almost the same as for PG1115+080 but is determined with a higher uncertainty. The result on $h$ is qualitatevely different; even if the two 90\% range do overlap enough, the histogram of $h$ values is peaked towards lower ones in clear contrast to what has been obtained for PG1115+080 and is expected. Kormann, Schneider \& Bartelmann (1994) have shown that the lensing observables for B1422+231 may be reproduced by an isothermal model ($\alpha = 1$) and an external shear and our result on $\alpha$ is in good agreement with their conclusion. This is encouraging since the lensing potential they use is the sume of the isothermal ellipsoidal model and and an external shear and thus belongs to the class of potentials described by Eqs.(\ref{eq: psigen}), (\ref{eq: psilens}) and (\ref{eq: psishear}) considered here. This could suggest that the model is not so bad, but this conclusion can not be enforced by a comparison with other estimates of $H_0$ simply because this is the first time that B1422+231 has been used to estimate the Hubble constant from time delays. However, we aware the reader that our estimate relies on the time delays measured by Patnaik \& Narasimha (2001) which are affected by quite high uncertainty. Choosing values for the time delays (consistent within the errors) different from the one we have used lead to a different time delay ratio and consequentely to different estimates of $\alpha$ and $h$. Given this situation, we have decided to still retain the results on $h$ coming from B1422+231 and combine them with the ones from PG115+080.

\subsection{Combining the results : final estimate of the Hubble constant}

Here we combine the histograms obtained for PG1115+080 and B1422+231 to get the final estimate of the Hubble constant. To this aim, we apply the same procedure used for the simulations and described in Sect. 5. The result we get is shown in Fig. 5 where we have used a binning of 0.1 for sake of clarity. Our final estimate for the Hubble constant is thus\,:

\begin{equation}
H_0 = 58_{-15}^{+17} \ {\rm km \ s^{-1} \ Mpc^{-1}} \ (68\% CL) \ ,
\label{eq: hzero68}
\end{equation} 
or more conservatively :
  
\begin{equation}
H_0 = 58_{-22}^{+26} \ {\rm km \ s^{-1} \ Mpc^{-1}} \ (90\% CL) \ .
\label{eq: hzero90}
\end{equation}  

Note that the error increases significantly when we consider the 90\% instead than the 68\% CL and this is essentially due to the presence of the peaks to lower values due to the unexpected results from B1422+231. Should this peak be eliminated (considering different time delays estimates for B1422+231), the 90\% range should be reduced too. However, also limiting ourserlves to the 68\% CL range, we note that our final estimate is in good agreement with the previous ones in literature. Williams \& Saha (2000) used the pixellated lens method to estimate $H_0$ from quadruple lenses combining the results from PG1115+080 and B1608+656 to finally get $H_0 = 61 \pm 11 \ {\rm km \ s^{-1} \ Mpc^{-1}}$ (68\% CL) which is in good agreement with our finding. The parametric technique has been applied to different lens systems, both double and quadruple. We have yet quoted the result obtained by Keeton \& Kochanek (1997) for PG1115+080; here we stress that our final estimate of $H_0$ is still in agreement with the value found by them. It is also interesting to compare the result we get with the one obtained by Koopmans \& Fassnacht (1999); applying a parametric technique to B1608+656 and considering different models, they finally quote $H_0 = 65^{+7}_{-6} \ {\rm km \ s^{-1} \ Mpc^{-1}}$ in satisfactory agreement with our 68\% CL on $H_0$. Finally, we also stress that our result is consistent also with completely different methods of estimate of the Hubble constant, such as the result of the HST Key Project, $H_0 = 72 \pm 8 \ {\rm km \ s^{-1} \ Mpc^{-1}}$ (\cite{HKP01}), using local estimators (as Cepheids and SNIa) and the one from an orientation unbiased sample of Sunyaev\,-\,Zeldovich clusters, $H_0 = 65^{+8}_{-7} \ {\rm km \ s^{-1} \ Mpc^{-1}}$ (\cite{J01}).

\section{Conclusions}

In this paper we have presented a new semianalytical method to estimate the Hubble constant from the measured time delays in quadruply imaged gravitational lens system. Assuming that the galaxy lens potential may be splitted into a radial part described by a simple power\,-\,law profile and an angular part described by a quite general shape function, we have been able to write down a nonlinear set of equations taking into account also the contribution of the external shear. This system may be solved numerically allowing us to fully take into account the lens models degeneracy; a set of physical constraints is then used to select the only reasonable solutions thus avoiding the risk of including in our considerations also non physically motivated models. The final set of solutions may be seen as a parametrization of that class of lens models which are able to generate the observed images configuration and the measured time delay ratio. The class of models so delineated may be translated in a sample of values for the Hubble constant $H_0$ which leads us to the final estimate of this cosmological parameter. 

To test the method we have applied it to three different simulated lens systems varying both the angular and the radial part of the lens potential and also the external shear parameters. This tests have shown us that the method indeed works and it is also very efficient in determining the slope $\alpha$ of the radial profile with a reasonable accuracy. Even if this were not our final aim, the ability of our method to find out the $\alpha$ parameter is a very interesting byproduct since this quantity is very usefull in modelling the dark halos which may be considered as the most important galaxy component responsable of the observed quadruply imaged lens system. The tests have also shown us that the method is not efficient in recovering the Hubble constant; even if the central value of the 90 \% range individuated for $H_0$ is very near to the input value, the range itself is quite large. This is not an unexpected result since we are fully taking into account the lens models degeneracy and it is well known that this increases the uncertainty on the Hubble constant. Motivated by this result, we have tried to reduce the uncertainty on $H_0$ combining the estimates from different systems. To this aim we simply build a combined histograms of the $H_0$ values (binned by $0.1 \ {\rm km \ s^{-1} \ Mpc^{-1}}$) multiplying the probabilities from each system and then excluding from the final sample those values of $H_0$ which turn out to have a final probability less than 1\%. This procedure allows us to reduce the uncertainty on $H_0$ by combining the three simulated systems. The final estimate of $H_0$ is perfectly consistent with the input value used for the simulations which confirms us that the method indeed works. 

Given these encouraging results, we have then applied our method to the real lenses PG1115+080 and B1422+231, which are two of the only three quadruple lenses for which the time delay has been measured. As regard PG1115+080, we find a slope $\alpha = 1.03_{-0.20}^{+0.24}$ (90\% CL) which indicates a near isothermal model consistent with previous models in literature. The 90\% range for $H_0$ turns out to be quite large ($25 \div 78 \ {\rm km \ s^{-1} \ Mpc^{-1}}$), but we note that the central value ($H_0 = 68 \ {\rm km \ s^{-1} \ Mpc^{-1}}$) is consistent with the values quoted in literature and obtained with different method. As regard B1422+231, we find $\alpha = 1.08_{-0.36}^{+0.33}$ and $H_0 = 25 \div 78 {\rm km \ s^{-1} \ Mpc^{-1}}$, with a distribution of values for the Hubble constant peaked towards lower ones. As we know, this is the first time that this system has been taken in consideration to determine the Hubble constant since the time delays have been measured only recently. Combining the resulting histograms from PG1115+080 and B1422+231 leads us to the following final estimate for the Hubble constant\,:

\begin{displaymath}
H_0 = 58_{-15}^{+17} \ {\rm km \ s^{-1} \ Mpc^{-1}} \ (68\% CL) \ ,
\end{displaymath} 
or more conservatively :
  
\begin{displaymath}
H_0 = 56_{-26}^{+22} \ {\rm km \ s^{-1} \ Mpc^{-1}} \ (90\% CL) \ .
\end{displaymath}  
 
We note that the uncertainty singnificantly increases when passing from the 68\% to the 90\% range for the estimate of $H_0$ and that the magnitude of the increasement is higher than expected when compared with the result from our simulations. Analyzing in detail the data, however, it is easy to see that this strange behaviour is completely due to the histogram obtained for B1422+231, which predicts too many models with low values of $H_0$. A possible explanation of this strange behaviour may be connected to the very high uncertainty on the measured time delays which translates to an high uncertainty on the recovered parameters $\alpha$ and $h$. Should we have chosen different values for the time delays (still within the uncertainties), we should have obtained lower values for $\alpha$ and higher values for $H_0$ thus narrowing the 90\% range. Some tests have suggested us that this could be actually a possible explanation. However, also considering only the 68\% CL, our final estimate of $H_0$ is consistent with all the previous estimates in literature whatever is the method used and the estimators considered.

{The new semianalytical method presented here turns out to be a valid complementary alternative to the usual parametric approach and to the fully non\,-\,parametric techniques. Parametric methods are usefull in building up galaxy models which may be easily compared to other galaxy observables (if possible) also not directly connected with the lensing charachteristics. However, they have only a modest power in exploring the parameter space since one cannot include too many parameters in the model to avoid having a number of degrees of freedom too low in the $\chi^2$ minimization. Given that the number of constraints is fixed (eight from images positions and two from time delays ratios if the system is a quadruple one), there in an unavoidable limit to the accuracy of the galaxy model and to the possibility to explore the wide range of lens models that may fit the same lensing observables. On the other hand, non parametric methods try to introduce as less {\it a priori} hypotheses as possible thus fully exploring the space of the models. However, a fully non parametric approach as the one adopted in the pixellated lens method of Williams \& Saha (2000) does not allow to control if the models considered are physically motivated or not thus risking to overestimate the uncertainties connected to the modelling problem. Our new method is less general than the pixellated one, but it has the advantage that it selects only models which are physically reliable. Besides, combining the results from different quadruple lens systems helps in reducing the other unidentified sources of systematic errors connected to single systems leading to a final estimate for $H_0$ which correctly takes into account all the possible sources of errors. 

Further improvements are however still possible. Numerical simulations of galaxy formation in different cosmological backgrounds have suggested that the dark halo density profile may be described by a simple universal law (see, e.g., \cite{NFW97}, \cite{Moore}, \cite{Klypin}). These models lead to lensing potentials which may not be described by Eq.(\ref{eq: psilens}) since the radial profile is not a single power\,-\,law, having different radial slopes for the inner and the outer parts. This could suggest that the potential we have used is not realistic and thus our results on the Hubble constant are wrong. However, we have shown that our estimates turn out to be in good agreement with all the previous ones obtained both with lensing based method and other distance ladders (such as Cepheid or SnIa). It is worth to note that the lensing observables are mainly determined by the mass distribution in the outer parts of the dark halos and this latter may be well described by models with a single power\,-\,law radial profile. Our models differ from the one predicted by numerical simulations only in the inner parts which are less important in determining the images positions and the time delays. It is thus expected that the difference does not introduce any serious systematic error. Anyway this does not mean that the results from numerical simulations are wrong since the statistic on the lensing systems considered is too low. To understand whether these models are really able to reproduce the lensing observables (number and position of images and time delay ratios) in quadruply imaged systems and whether they introduce a significative change in the estimate of $H_0$, one has to wait for more quadruple lenses with measured time delays. At the same time, it should be interesting to generalize our method in order to allow also a varying slope $\alpha$ of the radial profile thus possibly leading also to some constraints from the observed quadruply imaged QSOs.

Finally we note that the number of observed quadruple systems to which our method may be applied is considerable so that we have just to wait for the measure of the time delays to finally get an accurate estimate of the Hubble constant competitive with the ones coming from local estimators.

\begin{acknowledgements}

We warmly thank Sante Carloni and Olaf Wucknitz for the interesting discussions and suggestions.  
\end{acknowledgements}

\end{document}